\documentclass[prb,twocolumn,floatfix,showpacs,showkeywords]{revtex4}

\usepackage{amssymb}
\usepackage{amsmath}
\usepackage{amsfonts}
\usepackage{bm}
\usepackage{graphicx}

\renewcommand{\S}{{\mathcal{S}}}
\newcommand{\D}{{\mathcal{D}}}
\newcommand{\Ham}{{\hat{\mathcal{H}}}}
\newcommand{\Z}{{\mathcal{Z}}}
\newcommand{\G}{{\mathcal{G}}}
\newcommand{\C}{{\mathcal{C}}}

\newcommand{\bk}{{\bm k}}
\newcommand{\bx}{{\bm x}}

\newcommand{\br}{{\bm r}}

\newcommand{\hc}{{\textrm{h.c.}}}
\newcommand{\ud}{{\textrm{d}}}

\newcommand{\bhi}{{\hat{b}_i}}
\newcommand{\bhj}{{\hat{b}_j}}
\newcommand{\bhdi}{{\hat{b}^\dagger_i}}

\newcommand{\bhia}{{\hat{b}_{ia}}}
\newcommand{\bhdia}{{\hat{b}^\dagger_{ia}}}

\newcommand{\psiia}{{\psi^{a}_i}}
\newcommand{\psija}{{\psi^{a}_j}}
\newcommand{\psidia}{{\psi^{a*}_i}}

\newcommand{\phiia}{{\phi^{a}_i}}
\newcommand{\phija}{{\phi^{a}_j}}
\newcommand{\phidia}{{\phi^{a*}_i}}

\newcommand{\phiib}{{\phi^{b}_i}}
\newcommand{\phijb}{{\phi^{b}_j}}

\newcommand{\nia}{{\hat{n}_i^a}}
\newcommand{\nib}{{\hat{n}_i}^b}

\newcommand{\nn}{{\nonumber}}

\begin{document}

\title{Anomalous suppression of the Bose glass at commensurate fillings in the disordered Bose-Hubbard model}
\author{Frank Kr\"uger}
\author{Jiansheng Wu}
\author{Philip Phillips}
\affiliation{Department of Physics, University of Illinois, 1110 W. Green
St., Urbana, IL 61801, USA}

\date{\today}

\begin{abstract}
We study the weakly disordered Bose-Hubbard model on a cubic lattice through a one-loop renormalization group analysis
of the corresponding effective field theory which is explicitly derived by combining a strong-coupling expansion with a replica
average over the disorder. The method is applied not only to generic uncorrelated on-site disorder but also to simultaneous  
hopping disorder correlated with the differences of adjacent disorder potentials. Such correlations are inherent in fine-grained optical 
speckle potentials used as a source of disorder in optical lattice experiments. As a result of strong coupling, the strength of the replica mixing 
disorder vertex, responsible for the emergence of a Bose glass, crucially depends on the chemical potential and the Hubbard repulsion and vanishes to 
leading order in the disorder at commensurate boson fillings. As a consequence, at such fillings a direct transition between the 
Mott-insulator and the superfluid in the presence of disorder cannot be excluded on the basis of a one-loop calculation.
At incommensurate fillings, at a certain length scale, the Mott insulator will eventually become unstable towards 
the formation of a Boss glass. Phase diagrams as a function of the microscopic parameters are presented and the finite-size crossover between 
the Mott-insulating state and the Bose glass is analyzed. 
\end{abstract}

\maketitle

\section{Introduction}

The disordered Bose-Hubbard (BH) model can be viewed as the fruit fly for studying the interplay between disorder and strong interaction physics.  
In the absence of disorder, two phases compete, both of which have a gap in the single-paricle spectrum and hence are incompressible: the Mott 
insulator (MI) which wins when the on-site repulsions dominate and a superfluid (SF) which obtains when the kinetic energy exceeds a critical threshold. 
Disorder introduces a new avenue for localization independent of the interaction strength.  Fisher, et al.\cite{Fisher+89} argued that a Bose glass (BG), 
a gapless insulating state, always intervenes upon the disruption of the SF in the presence of disorder.  Nonetheless,  this finding remains controversial 
as simulations and analytical arguments both support \cite{Mukhopadhyay+96,Freericks+96,Svistunov96,Herbut97,Herbut98,Prokofev+04,Weichman+08,Pollet+09} 
and negate \cite{Scalettar+91,Krauth+91,Singh+92,Pai+96,Kisker+97,Pazmandi+98,Sen+01,Lee+01,Wu+08,Bissbort+09} this claim, the latter 
focussing predominantly on the case of commensurate boson fillings. In the absence of disorder, the suppression of density fluctuations at commensurate fillings 
is known to change the nature of the transitions from mean-field like to the universality class of the $(D+1)$-dimensional XY model\cite{Doniach81,Fisher+88} with $D$ the spatial dimension, suggesting that the system might respond differently to an infinitesimal amount of disorder. It has been argued by Fisher, 
et al.\cite{Fisher+89} that such a scenario is in principle possible but unlikely since disorder is expected to destroy commensuration. 

Certainly, the experimental realization of the MI-SF transition in a gas of ultracold bosonic atoms in an optical lattice\cite{Greiner+02} and the possibility of
introducing disorder in a highly controlled
manner,\cite{Lewenstein+07} e.g. by superimposing an optical speckle
potential,\cite{Lye+05,Clement+08,Billy+08,White+09}  has triggered
hope that experiments might address the questions raised above.
However, the experiments suffer from serious limitations, such as,
inhomogeneous densities due to the presence of an optical trap potential, heat generation by the 
lasers, and small system sizes, which might ultimately render them
incapable of answering questions about uniform systems in
the thermodynamic limit. This raises the question as to whether or not
optical lattice experiments will live up to the expectation that they provide the ultimate 
quantum simulators of paradigmatic Hamiltonians as the one of the disordered BH model, 

\begin{eqnarray}
\Ham & = & -\sum_{\langle i,j\rangle}t_{ij}(\bhdi\bhj+\hc)+\sum_i(\epsilon_i-\mu)\hat{n}_i\nn\\
& & +\frac{U}{2}\sum_i\hat{n}_i(\hat{n}_i-1).
\label{BHmodel}
\end{eqnarray}
This model describes bosons with corresponding creation (annihilation) operators
$\bhdi$ ($\bhi$) hopping with (disordered) amplitudes $t_{ij}$ between
nearest-neighbor bonds $\langle i,j\rangle$ of a $D$ dimensional
hypercubic lattice in the presence of an on-site disorder potential
$\epsilon_i$  and subject to a local Coulomb repulsion $U$. Further,
$\hat{n}_i=\bhdi\bhi$ denotes the occupation number operator and $\mu$
the chemical potential. In the following we revisit the general
disorder problem with an eye for the special form of disorder that is induced 
by a fine-grained optical speckle lens used in recent experiments.\cite{White+09} 

Detailed measurements of the speckle potential\cite{White+09} and subsequent calculations based on the imaginary time evolution of localized Wannier 
states in the presence of the measured potential\cite{White+09,Zhou+09} have
revealed the induced distributions of the microscopic parameters.  The 
potential disorder $\epsilon_i$ has been shown to follow the asymmetric distribution $P(\epsilon_i) =e^{-\epsilon_i/\Delta}/\Delta$ for $\epsilon_i\ge 0$ and 
$P(\epsilon_i) =0$ for $\epsilon_i< 0$. The reason for the one-sidedness of the distribution is the use of a blue-detuned speckle potential which can only increase 
the potential energy of a lattice site. Since correlations between
different sites are small even for nearest neighbor sites, we will assume the potential 
disorder to be uncorrelated. Simultaneously, the speckle field induces disorder in the hopping amplitudes which is correlated with the difference of the disorder potentials
on adjacent sites such that $t_{ij}=t+\gamma(\epsilon_i-\epsilon_j)^2$. Here, $t$ denotes the nearest-neigbor hopping in the absence of disorder. The relative width 
of the distribution of the on-site repulsion $U$ has been shown to be negligible. 

In this work we augment a strong-coupling expansion around the localized limit\cite{Sachdev99,Sengupta+05} with a replica average
over the disorder. In the clean limit, this procedure captures the quantum phase transition between the MI and the SF.  The resulting effective field theory is then 
studied within a one-loop renormalization group analysis.  The combination of strong-coupling expansion and replica calculus outlined in this work allows for a 
systematic calculation of thermodynamic phase diagrams of the disordered BH model. 

Our major finding is that  the effective replica-mixing disorder vertex, which is responsible for the emergence of a BG, depends crucially on the value of 
the chemical potential. Irrespective of the special form of the disorder, on approaching commensurate fillings, the disorder vertex is strongly suppressed and vanishes 
to leading order given by the variance of the disorder distribution. As a consequence, the length scale at which the MI becomes unstable towards the formation of a BG
becomes increasingly large. On the basis of a one-loop renormalization-group analysis which consistently treats the leading disorder contributions a direct MI-SF
transition remains possible in the presence of disorder at commensurate fillings. 
Whereas, as we show, a conclusive answer necessitates a considerably more complicated 2-loop calculation, the anomalous
suppression of the generic disorder vertex and the presence of additional disorder operators which turn marginal at commensurate fillings strongly suggests that 
the nature of the transition at commensuration is in a different universality class from the transition at  incommensurate fillings.

The outline of the paper is as follows. In Sec. \ref{sec.effective},
we derive the effective field theory for the disordered BH model in
the strong-coupling and weak disorder regime, providing explicit
expressions relating the effective coupling constants to the
microscopic parameters. To lay plain the details of the derivation, we start with a brief review of the Hubbard-Stratonovich transformation underlying the strong-coupling 
expansion in the clean limit (Sec. \ref{subsec.clean}) and illustrate the nontrivial coupling of on-site potential disorder to 
the dual theory. Employing the replica trick to average over the
disorder, we obtain the effective field theory (Sec. \ref{subsec.onsite}). In the case of simultaneous 
hopping and on-site disorder, we derive the effective field theory by first performing a disorder average of the replicated system and by then generalizing the consecutive 
Hubbard-Stratonovich transformation and strong-coupling expansion
(Sec. \ref{subsec.speckle}). Starting from the effective action, we derive in Sec. \ref{sec.rg} the one-loop 
renormalization-group equations.  We present the integrated numerical
solutions in Sec. \ref{subsec.numint} which enable a direct study of
the instability of the MI towards the formation of a BG. The resulting
phase diagrams are presented in Sec. \ref{sec.results}. Finally, in
Sec. \ref{sec.disc}, we summarize our results, compare with other
analytical work, and discuss the relevance of our findings  to recent optical lattice experiments.

\section{Effective field theory}
\label{sec.effective}

In this section, we derive the effective long-wavelength field theory for the weakly disordered BH model in the strong-coupling limit. We focus on temperatures much 
smaller than the on-site Hubbard repulsion, $\beta U\ll 1$ where $\beta$ denotes the inverse temperature. In order to set up a field theoretical description, it is useful to
express the partition function of the disordered BH model
(\ref{BHmodel}) as a path integral in imaginary time $\tau\in[0;\beta)$  over complex bosonic coherent states,
$\bhi |\phi_i\rangle=\phi_i|\phi_i\rangle$, 

\begin{eqnarray}
\Z & = & \int\D[\phi,\phi^*]e^{-\S_\phi}\nn\\
\S_\phi & = & \int_0^\beta\ud\tau\left\{\sum_i\left(\phi_i^*\partial_\tau\phi_i+(\epsilon_i-\mu)|\phi_i|^2+\frac{U}{2}|\phi_i|^4  \right)\right.\nn\\
            & & \left. -\sum_{\langle ij\rangle}[t+\gamma(\epsilon_i-\epsilon_j)^2](\phi_i^*\phi_j+\phi_i\phi_j^*) \right\},
\label{BHPI}
\end{eqnarray}
where we have explicitly built in the correlation between the on-site disorder potential $\epsilon_i$ and the hopping disorder $\delta t_{ij}$. Note that in this model
a perfect correlation between the on-site and hopping disorder is assumed. This seems reasonable for disorder induced by an optical speckle lens since all distributions 
of the microscopic parameters as well as the correlations between them originate from the same speckle disorder potential. A possible generalization would by a model 
with two separate distributions $P(\epsilon_i)$ and $P(\delta t_{ij})$ with tunable cross-correlations.

\subsection{Clean system}
\label{subsec.clean}

We start with a brief review of the derivation of the effective strong-coupling theory in the clean system ($\epsilon_i=0$ on all lattice sites). The formation of the MI state 
takes place in the regime where the Coulomb repulsion $U$ dominates over the hopping amplitude $t$. Therefore, it is desirable to find a dual description in which residual 
vertex corrections are controlled by the smallness of $t/U$. Such an expansion around the localized limit can be achieved by a Hubbard-Stratonovich transformation as
sketched in Refs. [\onlinecite{Fisher+89,Sachdev99}] and outlined in detail in Ref. [\onlinecite{Sengupta+05}]. The crucial step consists in the decoupling of the boson hopping 
term by introducing a complex auxiliary field $\psi_i(\tau)$, 

\begin{eqnarray}
\Z & = & \int\D[\phi,\phi^*,\psi,\psi^*]e^{-(\S_\phi^{(0)}+\S_\psi^{(0)}+\S_{\phi\psi})},\nn\\
\S_\phi^{(0)} & = &  \int_0^\beta\ud\tau\sum_i\left(\phi_i^*\partial_\tau\phi_i-\mu|\phi_i|^2+\frac{U}{2}|\phi_i|^4  \right),\nn\\
\S_\psi^{(0)} & = &  \int_0^\beta\ud\tau\sum_{ij}\psi_i^*(T^{-1})_{ij}\psi_j,\nn\\
\S_{\phi\psi} & = &  \int_0^\beta\ud\tau\sum_i\left(\phi_i^*\psi_i+\phi_i\psi_i^*\right),
\end{eqnarray}
which is easily verified by completing the square and integrating over the fields $\psi_i(\tau)$. Here, $T^{-1}$ denotes the inverse of the hopping matrix which
is given by $t_{ij}=t$ for $i,j$ nearest neighbors and $t_{ij}=0$
otherwise. After taking the trace over the original bosonic fields
$\phi_i(\tau)$, we rewrite the partition function of the
BH model as 

\begin{eqnarray}
\Z & = & \Z_0\int\D[\psi,\psi^*]e^{-(\S_\psi^{(0)}+\S_\psi')}\nn\\
\S_\psi' & = &  -\ln\langle T_\tau\exp[ \int_0^\beta\ud\tau\sum_i(\psi_i(\tau)\bhdi(\tau)+\hc)] \rangle_0,\quad
\label{cumulant}
\end{eqnarray}
where the average $\langle\ldots\rangle_0$ has to be taken with respect to the local on-site Hamiltonian
$\Ham_0 = \sum_i[\frac{U}{2} \hat{n}_i(\hat{n}_i-1)-\mu \hat{n}_i]$ which is diagonal in the local occupation number basis $|\{n_i\}\rangle$. Further, $\Z_0$ denotes the 
partition function in the localized limit, $\Z_0=\textrm{Tr}e^{-\beta\Ham_0}$. Note that we re-expressed the average over bosonic fields $\phi_i(\tau)$ as an operator average
where $\bhdi(\tau)=\exp(\Ham_0\tau)\bhdi \exp(-\Ham_0\tau)$ and $T_\tau$, the time-ordering operator. 

In principle, the part $\S_\psi'$ can be expanded to any desired order in the fields $\psi$ since the coefficients are simply related to bosonic Green functions of the local 
Hamiltonian $\Ham_0$. However, already the calculation of the coefficients of the quartic terms $\sim\psi^4$ is tedious since it necessitates the evaluation of the two-particle 
bosonic Green function. Note that $\Ham_0$ represents an \emph{interacting} local problem and consequently Wick's theorem does not apply.
After expanding $\S_\psi'$ up to quartic order, we take the continuum limit and perform a temporal and spatial gradient expansion to obtain the effective action

\begin{eqnarray}
\S_\textrm{eff} & = & \int_0^\beta\ud\tau\int\ud^D\br\left(K_1^{(0)} \psi^* \partial_\tau \psi+K_2^{(0)} |\partial_\tau \psi|^2\right.\nn\\
& & \left.+K_3^{(0)} |\nabla  \psi|^2+R^{(0)}|\psi|^2+H^{(0)}|\psi|^4\right).
\label{effactclean}
\end{eqnarray}
Here, the spatial gradient term with coefficient $K_3^{(0)}=1/(zt)$ results from the long-wavelength expansion of $\S_\psi^{(0)}$ and $z=2D$ denotes the coordination
number of the $D$-dimensional hypercubic lattice. The mass coefficient of the theory has contributions from both $\S_\psi^{(0)}$ and $\S_\psi'$ and is given by
$R^{(0)}=K_3^{(0)}+\G^{(0)}$ where

\begin{equation}
\G^{(0)}=-\int_{-\beta}^\beta\ud\tau\langle T_\tau\bhi(\tau)\bhdi(0)\rangle_0
\end{equation}
is identical to the Fourier transform of the single particle Green function $G_i(\tau-\tau')=-\langle T_\tau\bhi(\tau)\bhdi(\tau')\rangle_0$ in the zero-frequency limit. 
In order to calculate $G_i(\tau)$, we first realize that at sufficiently low temperatures ($\beta U\ll1$) the partition function is dominated by the ground-state configuration 
$n_i=m$ where $m$ is the integer minimizing the site energy $\epsilon_n=-\mu n+\frac{U}{2}n(n-1)$, hence the smallest integer larger than $\mu/U$. In this approximation 
the single-particle Green function is easily calculated by inserting a complete set of states, $1=\sum_n |n\rangle\langle n|$, 

\begin{equation}
G_i(\tau) = -(m+1)e^{-\epsilon_+\tau}\Theta(\tau)-m e^{\epsilon_-\tau}\Theta(-\tau).
\end{equation}
Here $\Theta(x)=1$ for $x>0$ and $\Theta(x)=0$ for $x<0$ denotes the Heavyside 
function. For brevity, we have defined $\epsilon_+=\epsilon_{m+1}-\epsilon_m=mU-\mu$
and $\epsilon_-=\epsilon_{m-1}-\epsilon_m=(1-m)U+\mu$. Using this result, we obtain the mass coefficient in the zero-temperature limit,

\begin{equation}
R^{(0)}=\frac{1}{zt}-\left(\frac{m+1}{\epsilon_+}+\frac{m}{\epsilon_-}\right).
\end{equation}
The coefficients of the temporal gradient terms are given by derivatives of the mass coefficient with respect to the chemical potential, 
$K_1^{(0)}=-\frac{\partial R^{(0)}}{\partial\mu}$ and $K_2^{(0)}=-\frac 12\frac{\partial^2R^{(0)}}{\partial\mu^2}$, which directly follows from the temporal 
gradient expansion but can also be shown to be a consequence of the $U(1)$ gauge symmetry.\cite{Sachdev99}

The interaction vertex is given by the connected parts of the two particle Green function in the static limit. In order to calculate the two particle Green function,
$G_i^{\textrm{II}}(\tau_1,\tau_2,\tau_3,\tau_4)=\langle T_\tau \bhi(\tau_1) \bhi(\tau_2) \bhdi(\tau_3) \bhdi(\tau_4)  \rangle$, for every possible time ordering we simply insert identities
$1=\sum_n |n\rangle\langle n|$ and evaluate the resulting products of matrix
elements of bosonic creation and annihilation operators. The calculation
is tedious but straightforward.\cite{Sengupta+05} Here we provide only the final result which is given by 

\begin{eqnarray}
H^{(0)} & = & \left(\frac{m}{\epsilon_-}+\frac{m+1}{\epsilon_+} \right)\left(\frac{m}{\epsilon_-^2}+\frac{m+1}{\epsilon_+^2} \right)\nn\\
& & -\frac{m(m-1)}{\epsilon_-^2\epsilon_{-2}} -\frac{(m+1)(m+2)}{\epsilon_+^2\epsilon_{+2}}
\end{eqnarray}
with $\epsilon_\pm$ as defined above, $\epsilon_{+2}=\epsilon_{m+2}-\epsilon_m=(1+2m)U-2\mu$, and 
$\epsilon_{-2}=\epsilon_{m-2}-\epsilon_m=(3-2m)U+2\mu$. 

To summarize, we have derived an effective strong-coupling field theory providing the explicit dependence of the effective parameters on the
microscopic ones, namely the on-site Coulomb repulsion $U$, the hopping $t$, and the chemical potential $\mu$. Note that in this dual representation, 
the vertex correction $H^{(0)}$ is of relative order $t/U$ as a consequence of the underlying expansion around the localized limit. We have neglected vertex 
corrections containing additional temporal derivatives.

\begin{figure}[t]
\includegraphics[width=\linewidth]{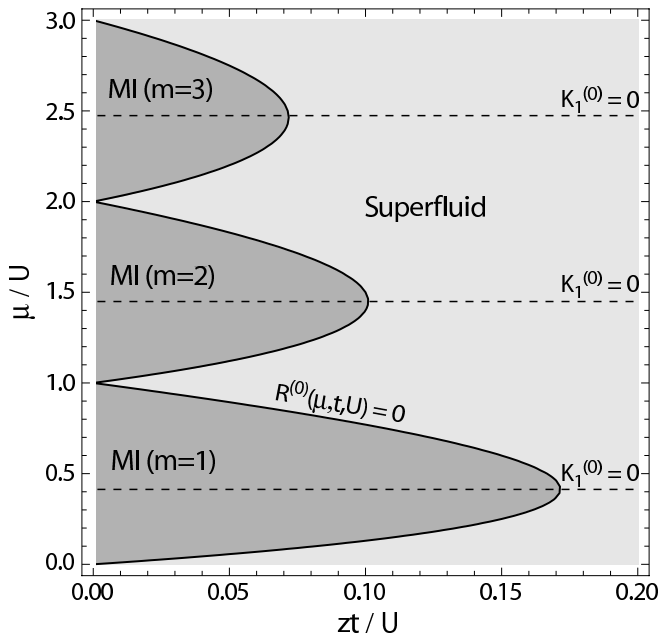}
\caption{Mean-field phase diagram of the clean BH model as a function of $\mu/U$ and $t/U$ showing the first three Mott-lobes with $m=1,2,3$ bosons
per lattice site. The phase boundaries are obtained by the sign change of the mass coefficient $R^{(0)}(\mu,t,U)$. The dashed lines indicate the tip positions of the lobes at which
the linear time derivative term in the effective field theory vanishes, $K_1^{(0)}=0$.}
\label{fig.clean}
\end{figure}

At mean-field, the phase boundaries between the MI states with fillings $m$ and the SF are obtained by the sign change of the mass coefficient $R^{(0)}(\mu,t,U)$.
The SF corresponds to the broken-symmetry phase ($R^{(0)}<0$) where the SF density is given by $\rho_s\sim\langle \hat{b}\rangle=K_3^{(0)}\langle\psi\rangle$.\cite{Sachdev99}
The MI states are obtained for $R^{(0)}>0$ where the mass coefficient corresponds to the Mott gap. 
We show in Fig. \ref{fig.clean} the mean-field phase diagram as a function of the dimensionless parameters $\mu/U$ and $t/U$. The tips of the different Mott lobes at
$(\mu/U)_m=\sqrt{m(m+1)}-1$ correspond to commensurate boson fillings
which greatly stabilize the MI states. 

Note that the coefficients $K_1^{(0)}$ and $K_2^{(0)}$
are given by first and second derivatives of the mass coefficient $R^{(0)}$ with respect to the chemical potential and are therefore related to the slope and curvature of the
phase boundaries. At the tips, we find $K_1^{(0)}=0$ and  $K_2^{(0)}>0$ yielding a field theory with dynamical exponent $d_z=1$, whereas at incommensurate fillings  
$K_1^{(0)}\neq0$ and $d_z=2$. From simple power counting, we find that at incommensurate fillings, the interaction vertex is an irrelevant perturbation for $D+2>4$ where $D$ 
denotes the spatial dimension and therefore, for $D=3$ the transitions
are mean field. However, despite being irrelevant, the interaction vertex leads to a renormalization
of the mass coefficient and therefore to a slight shift of the phase boundaries as we will see in Sec. \ref{sec.results}.

\subsection{Pure on-site disorder}
\label{subsec.onsite}

As a next step, we use the results of Sec. \ref{subsec.clean} to derive the effective field theory in the case of weak on-site disorder $\epsilon_i$ with a probability 
distribution $P(\epsilon_i)$. Without loss of generality we can assume that the distribution has zero mean, $\overline{\epsilon}=\int \ud\epsilon P(\epsilon)\epsilon=0$ and 
variance $\overline{\epsilon^2}=\int \ud\epsilon P(\epsilon)\epsilon^2
=\Delta^2$. In case of distributions with non-zero mean such as that induced by an optical speckle field, we simply shift the distribution and redefine the chemical potential as $\overline{\mu}=\mu-\overline{\epsilon}$. In the following we focus on weak disorder, 
$\Delta/U\ll 1$ and therefore neglect contributions from higher moments of the distribution. 
Further, we assume the disorder to be uncorrelated on different sites which is a good approximation in the case of fine-grained speckle disorder.\cite{White+09,Zhou+09}

Since disorder enters only in the form of random shifts of on-site
energies, the Hubbard-Stratonovich transformation outlined in the previous section \ref{subsec.clean} can be performed in 
exactly the same way, yielding the action 

\begin{eqnarray}
\S_\textrm{dis} & = & \int_0^\beta\ud\tau\left\{ \sum_{ij}\psi^*_i(T^{-1})_{ij}\psi_j+\sum_i \left(K_{1i} \psi_i^* \partial_\tau \psi_i\right.\right.\nn\\
& & \left.\left.\phantom{ \int_0^\beta}+K_{2i} |\partial_\tau \psi_i|^2 + \G_i|\psi_i|^2+H_i |\psi_i|^4\right)\right\},
\label{effactdis}
\end{eqnarray}
where the coefficients are given by the ones of the clean system but
with the chemical potential shifted by the disorder potential
$\epsilon_i$ on the corresponding site.  The contribution to mass term takes the form

\begin{eqnarray}
\G_i & = &  \G^{(0)}(\mu-\epsilon_i,U)\nn\\
& = &  -\left(\frac{m}{(1-m)U+\mu-\epsilon_i}+\frac{m+1}{mU-\mu+\epsilon_i}\right).
\end{eqnarray}
From the above equation it is clear that the coupling to disorder is radically different from a generic 
random mass problem where one assumes disorder to couple linearly to the mass of the theory. Due to the strong-coupling expansion, the effective mass setting the Mott 
gap depends in a complicated way on the chemical potential of the BH model and therefore on the disorder.

In order to restore translational symmetry and to perform a continuum limit, we employ the standard replica trick to average over the disorder. To calculate the disorder-averaged free energy, one has to average the logarithm of the partition function, $\ln\Z=\lim_{n\to 0}(\Z^n-1)/n$. Therefore, we introduce
$n$ replicas of the system, $\psiia(\tau)$ with $a=1,\ldots,n$, the replica index, and define the effective, disorder averaged action $\S_\textrm{eff}$ as

\begin{eqnarray}
\overline{\Z^n} & = &  \int\D[\psiia,\psiia^*]\overline{e^{-\sum_a \S_\textrm{dis}[\psiia,\psiia^*]}}\nn\\
& =: &  \int\D[\psiia,\psiia^*]e^{-\S_\textrm{eff}[\{\psiia,\psiia^*\}]},
\end{eqnarray}
where $\overline{(\cdots)}$ denotes the average over the
disorder. The cumulant expansion through second order can be written compactly

\begin{equation}
\S_\textrm{eff} =  \sum_a \overline{\S_\textrm{dis}^a}-\frac 12 \sum_{ab}\left( \overline{\S_\textrm{dis}^a\S_\textrm{dis}^b}- \overline{\S_\textrm{dis}^a}\phantom{.}\overline{\S_\textrm{dis}^b}\right),
\end{equation}
in terms of  $\S_\textrm{dis}^a=\S_\textrm{dis}[\psiia,\psiia^*]$. 
After taking the continuum limit and performing the spatial gradient
expansion, we find that
\begin{eqnarray}
\S_\textrm{eff} & = & \int_0^\beta\ud\tau\int\ud^D\br \sum_a\left(K_1 \psi_a^* \partial_\tau \psi_a+K_2 |\partial_\tau \psi_a|^2\right.\nn\\
& & \left.+K_3 |\nabla  \psi_a|^2+R|\psi_a|^2+H|\psi_a|^4\right)\nn\\
& & +G\sum_{ab}\int_{\tau\tau'}\int\ud^D r \left|\psi_a(\tau)\right|^2|\psi_b(\tau')|^2,
\label{effact}
\end{eqnarray}
where we have again neglected all vertex corrections containing
imaginary-time derivatives. The effects of the potential disorder are
twofold. First, the mass and interaction vertices are renormalized as

\begin{eqnarray}
R & = & K_3^{(0)}+\overline{\G_i} = R^{(0)}-K_2^{(0)}\Delta^2,\\
H & = &  \overline{H_i} = H^{(0)}+\frac 12 \frac{\partial^2 H^{(0)}}{\partial\mu^2}\Delta^2,
\end{eqnarray}
respectively. As before, the coefficients of the temporal gradient terms are given by derivatives of the mass coefficient with respect 
to the chemical potential, $K_1=-\frac{\partial R}{\partial\mu}$ and $K_2=-\frac 12\frac{\partial^2R}{\partial\mu^2}$. Since the disorder does not couple to the 
hopping-matrix elements, the spatial gradient term remains unchanged, $K_3=K_3^{(0)}=1/(zt)$. Second, disorder generates a vertex correction which mixes different replicas
and is non-local in imaginary time, generic for effective disorder terms on the level of the replica averaged effective action. The strength of the disorder vertex is

\begin{equation}
G = -\frac 12 \left(\overline{\G_i^2}-\overline{\G_i}^2 \right) = -\frac 12 \left(K_1^{(0)}\right)^2\Delta^2.
\end{equation}
As it should, in the clean limit ($\Delta=0$) the replica mixing term is absent  and the effective action (\ref{effact}) reduces to $n$ identical copies of that for the clean system.
Note that the effective disorder strength $G$ is not simply given by
the variance $\Delta^2$ of the microscopic disorder but is also a function of the microscopic parameters
$\mu$ and $U$ entering via the factor $K_1^{(0)}$. As we have seen
above, this 
is a consequence of the underlying strong-coupling expansion and the resulting non trivial 
coupling to disorder. As we have realized before, $K_1^{(0)}=-\frac{\partial R^{(0)}}{\partial\mu}$ has a simple geometrical meaning and is related to the slope of the 
phase boundaries between the MI and superfluid states. To be more precise, $K_1^{(0)}$ is the projection of the gradient of $R^{(0)}$ on the chemical potential axis. Since
the gradient is perpendicular to the lines of constant mass, that is,
the mean-field phase boundary given by $R^{(0)}=0$, $K_1^{(0)}$
changes sign at the tip 
positions 
$(\mu/U)_m=\sqrt{m(m+1)}-1$ of the Mott lobes. As a consequence, the disorder vertex $G$ vanishes at such commensurate fillings.

\subsection{Speckle disorder}
\label{subsec.speckle}

In this section, we proceed to derive the effective field theory for the BH model in the presence of weak speckle disorder.  We assume a one-sided distribution 
$P(\epsilon_i) =e^{-\epsilon_i/\Delta}/\Delta$ for $\epsilon_i\ge 0$ and $P(\epsilon_i) =0$ for $\epsilon_i< 0$ of the disorder
potentials as extracted for optical speckle disorder\cite{Zhou+09} and assume that the disorder potentials are uncorrelated on different sites.
Since the speckle potential not only introduces disorder in the on-site energies as studied in the previous section but also in the nearest-neighbor hopping amplitudes, 
$\delta t_{ij}=\gamma(\epsilon_i-\epsilon_j)^2$, it is not possible to decouple the boson hopping term before performing a disorder average. Such a procedure would require the inversion of a disordered hopping matrix. Alternatively, we can first perform a disorder average and then generalize the Hubbard-Stratonovich transformation to 
decouple terms which are not site-diagonal.

Starting from the coherent state path-integral representation (\ref{BHPI}) of the speckle disordered BH model, we again employ the replica trick to obtain the 
disorder-averaged action  

\begin{eqnarray}
\S_\phi & = & \S_\phi^{(0)}+\S_\phi^{(1)}+\S_\phi^{(2)},\\
\S_\phi^{(0)} & = & \sum_{ia}\int\ud\tau \left(\phidia\partial_\tau\phiia-\overline{\mu}|\phiia|^2+\frac{U}{2}|\phiia|^4  \right),\nn\\
\S_\phi^{(1)} & = & -\frac{\Delta^2}{2}\sum_{iab}\int\ud\tau\int\ud\tau' |\phiia(\tau)|^2 |\phiib(\tau')|^2,\nn\\
\S_\phi^{(2)} & = & -\overline{t}\sum_{\langle ij\rangle a}\int\ud\tau\left(\phidia\phija+\hc\right),\nn\\
        & & +2\gamma\Delta^3 \sum_{\langle ij\rangle ab}\int\ud\tau\int\ud\tau'\left(\phidia(\tau)\phija(\tau)+\hc\right),\nn\\
        & & \phantom{+2\gamma\Delta^3 \sum_{\langle ij\rangle ab}} \times\left(|\phiib(\tau')|^2+|\phijb(\tau')|^2 \right),\nn
\end{eqnarray}
where we have used the disorder averages $\overline{\epsilon_i\epsilon_j}=\Delta^2(1+\delta_{ij})$ and 
$\overline{(\epsilon_i-\epsilon_j)^2\epsilon_k}=2\Delta^3(1+\delta_{ik}+\delta_{jk})$ for $i\neq j$, which are easily derived from the moments $\overline{\epsilon^n}=n!\Delta^n$
since the potentials are uncorrelated on different sites. Further, we have defined the disorder-shifted chemical potential and hopping
as $\overline{\mu}=\mu-\overline{\epsilon_i}=\mu-\Delta$ and $\overline{t}=t+\gamma\overline{(\epsilon_i-\epsilon_j)^2} =t+2\gamma\Delta^2$, respectively. Here, $\S_\phi^{(1)}$
results from the on-site disorder. Since to order $\Delta^2$ the coupling of the disorder to the hopping term has only the trivial effect of shifting the nearest neighbor 
hopping amplitude, we have included the next order ($\Delta^3$) replica mixing term  which corresponds to nearest neighbor hopping coupled to the occupation on adjacent sites.

Despite the more complicated structure due to this contribution, it is still possible to decouple the boson hopping terms $\S_\phi^{(2)}$ by introducing auxiliary 
Hubbard-Stratonovich fields $\psiia(\tau)$ as

\begin{eqnarray}
e^{-\S_\phi^{(2)}} & = & \int\D[\psiia,\psidia] e^{-(\S_\psi^{(0)}+\S_{\phi\psi}^{(1)}+\S_{\phi\psi}^{(2)})},\\
\S_\psi^{(0)} & = & \sum_{ija}\int\ud\tau\left(\overline{T}^{-1} \right)_{ij} \psidia\psija,\nn\\
\S_{\phi\psi}^{(1)} & = & -\sum_{ia}\int\ud\tau \left(\psiia\phidia  +\hc\right),\nn\\
\S_{\phi\psi}^{(2)} & = & \frac{2\gamma\Delta^3}{\overline{t}}\sum_{iab}\int_{\tau\tau'} \left(\psiia(\tau)\phidia(\tau)|\phiib(\tau')|^2+\hc\right), \nn
\end{eqnarray} 
where $\overline{T}^{-1}$ denotes the inverse of the hopping matrix $\overline{T}$ which has non-zero elements $\overline{t}$ for nearest neighbors. After taking the trace
over the boson fields $\phiia(\tau)$, $\phidia(\tau)$, we write the effective action formally as

\begin{equation}
\S_\textrm{eff} = \S_\psi^{(0)}-\ln \left\langle e^{-(\S_\phi^{(1)}+\S_{\phi\psi}^{(1)}+\S_{\phi\psi}^{(2)})}  \right\rangle_0,
\end{equation}
where the average $\langle\ldots\rangle_0$ over the bosonic fields $\phiia(\tau)$ has to be taken with respect to the action $\S_\phi^{(0)}$ corresponding to the
site-diagonal replicated Hamiltonian $\Ham_0^{(n)}=\sum_{ia}[-\overline{\mu}\nia+\frac{1}{2}U\nia(\nia-1)]$. As for the clean system, the averages in the cumulant expansion
are conveniently calculated by using the correspondence with time-ordered products of boson operators $\bhdia(\tau)=\exp(\Ham_0^{(n)}\tau)\bhdia\exp(-\Ham_0^{(n)}\tau)$
and $\bhia(\tau)$. These can easily be evaluated by inserting a
complete set of bosonic states in the occupation number basis. The terms of order $\Delta^0$ are again related to the
one and two-body bosonic Green functions and the calculation is completely analogous to the one outlined in section \ref{subsec.clean}. 

Extra complications arise in the calculation of the disorder terms. The contributions of order $\Delta^2$ from the on-site disorder are given by averages containing an additional 
factor $\S_\phi^{(1)}$. The contribution to the bilinear action ($\sim\psi^2$) is given by $\frac 12 \langle \S_\phi^{(1)}(\S_{\phi\psi}^{(1)})^2 \rangle_0$, to the quartic terms by
$\frac{1}{4!} \langle \S_\phi^{(1)}(\S_{\phi\psi}^{(1)})^4 \rangle_0-\frac 14 \langle \S_\phi^{(1)}(\S_{\phi\psi}^{(1)})^2 \rangle_0 \langle(\S_{\phi\psi}^{(1)})^2 \rangle_0$.
Therefore, one has to insert two additional occupation number operators $\nia(\tau)$, $\nib(\tau')$ into the operator products making the calculation of the average more 
tedious. Likewise, the terms of order $\Delta^3$ contain exactly one factor $\S_{\phi\psi}^{(2)}$ requiring the insertion of one extra occupation number operator. Since the
calculation is lengthy but straightforward and not particularly insightful we will only give the final results.

After taking the continuum limit and performing a temporal and spatial
gradient expansion, we obtain an effective action identical to
Eq. \ref{effactdis} but with a mass coefficient
 
\begin{eqnarray}
R  & = &  \frac{1}{z\overline{t}}-\left[\frac{m}{\epsilon_-}+\frac{m+1}{\epsilon_+}+4\frac{\gamma\Delta^3}{\overline{t}}
\left(\frac{m}{\epsilon_-^2}-\frac{m+1}{\epsilon_+^2} \right) \right.\nn\\
& & +\left. \Delta^2\left(\frac{m}{\epsilon_-^3}+\frac{m+1}{\epsilon_+^3} \right) \right]\nn\\
& = & R^{(0)}+4K_1^{(0)}\frac{\gamma\Delta^3}{\overline{t}}-K_2^{(0)}\Delta^2,
\label{masscoeff}
\end{eqnarray}
where $R^{(0)}$, $K_1^{(0)}$, and $K_2^{(0)}$ are identical to the coefficients of the clean system but with the hopping amplitude and chemical potential replaced by the disorder 
averaged values, $\overline{t}=t+2\gamma\Delta^2$ and $\overline{\mu}=\mu-\Delta$, respectively. As before, $m$ labels the different Mott lobes, $K_1=-\frac{\partial R}{\partial\mu}$,  
$K_2=-\frac 12 \frac{\partial^2 R}{\partial\mu^2}$, and $K_3=1/(z\overline{t})$ with $z=2D$ the coordination number. The dependence of the coefficients on the 
chemical potential and the on-site Coulomb repulsion enters via the functions $\epsilon_-=(1-m)U+\overline{\mu}$ and $\epsilon_+=mU-\overline{\mu}$. For the 
interaction vertex, we find

\begin{eqnarray}
H & = & \left(\frac{m}{\epsilon_-}+\frac{m+1}{\epsilon_+} \right)\left(\frac{m}{\epsilon_-^2}+\frac{m+1}{\epsilon_+^2} \right)-\frac{m(m-1)}{\epsilon_-^2\epsilon_{-2}} \nn\\
& & -\frac{(m+1)(m+2)}{\epsilon_+^2\epsilon_{+2}}+8\frac{\gamma\Delta^3}{\overline{t}}\left[\sum_{k=0}^4\frac{c_k m(m+1)}{\epsilon_-^{4-k}\epsilon_+^k}\right.\nn\\
& & \left. -2\sum_{k=1}^2\left(\frac{m(m-1)}{\epsilon_-^{4-k}\epsilon_{-2}^k}-\frac{(m+1)(m+2)}{\epsilon_+^{4-k}\epsilon_{+2}^k}  \right)\right]\nn\\
& & +\Delta^2\left[\sum_{k=0}^5\frac{d_k m(m+1)}{\epsilon_-^{5-k}\epsilon_+^k} -\sum_{k=1}^3 e_k\left(\frac{m(m-1)}{\epsilon_-^{5-k}\epsilon_{-2}^k}\right. \right.\nn\\
& & \left.\left.+\frac{(m+1)(m+2)}{\epsilon_+^{5-k}\epsilon_{+2}^k}\right)\right]\nn\\
& = & H^{(0)}-8\frac{\partial H^{(0)}}{\partial\mu}\frac{\gamma\Delta^3}{\overline{t}}+\frac 12 \frac{\partial^2 H^{(0)}}{\partial\mu^2}\Delta^2,
\end{eqnarray}
where $c_0=3m/(m+1)$, $c_1=-c_3=2$, $c_2=0$, $c_4=-3(m+1)/m$, $d_0=2c_0$, $d_1=d_4=3$, $d_2=d_3=-1$, $d_5=-2c_4$, $e_1=3$, $e_2=e_3=4$, and
$\epsilon_{-2}=(3-2m)U+2\overline{\mu}$,
$\epsilon_{+2}=(1+2m)U-2\overline{\mu}$. Finally, the replica mixing
disorder vertex reduces to

\begin{eqnarray}
G   & = &  -4\frac{\gamma\Delta^3}{\overline{t}}\left(\frac{m}{\epsilon_-}+\frac{m+1}{\epsilon_+} \right)\left(\frac{m}{\epsilon_-^2}-\frac{m+1}{\epsilon_+^2} \right)\nn\\
& & -\frac 12  \left(\frac{m}{\epsilon_-^2}-\frac{m+1}{\epsilon_+^2} \right)^2\nn\\
& = &  4K_1^{(0)}\left(K_3^{(0)}-R^{(0)}\right)\frac{\gamma\Delta^3}{\overline{t}}-\frac 12 \left(K_1^{(0)}\right)^2\Delta^2.\quad
\label{dis}
\end{eqnarray}
In the absence of the speckle, $G=0$, and the effective action (\ref{effactdis}) simply corresponds to $n$ identical
copies of the action for the clean system as given in section \ref{subsec.clean} and as derived previously.\cite{Sengupta+05} Switching off the coupling of the disorder to 
the hopping matrix elements ($\gamma=0$), we recover the case of pure uncorrelated potential disorder. In fact, for $\gamma=0$, the coefficients derived 
in this section are found to be identical with the ones derived in
section \ref{subsec.onsite} demonstrating that our results are
independent of the order in which the Hubbard-Stratonovich transformation and the replica disorder average are implemented. Whereas for pure on-site disorder, it is more economical 
to start with the Hubbard-Stratonovich transformation, the presence of simultaneous off-diagonal disorder enforces us to first average over the disorder and to perform 
the more tedious strong-coupling expansion afterwards.   

Interestingly, even in the presence of simultaneous correlated hopping disorder ($\gamma>0$), the disorder vertex $G$ vanishes as $K_1^{(0)}=0$ determining the
positions of the Mott-lobe tips in the clean system. At least to order $\Delta^3$, at such commensurate fillings, the only effect of disorder 
is to renormalize the coefficients of the clean system suggesting that a direct MI to SF transition obtains in the presence of weak disorder.

\section{Renormalization Group}
\label{sec.rg}

To investigate the effects of weak disorder on the phase diagram of the BH model and in particular to study the instability of the MI state towards the 
formation of a BG, we analyze the effective field theory (\ref{effact}) by means of the renormalization group (RG) method. In this section, the one-loop RG
equations are derived. Further, it is explained how the different phases can be identified from the RG flow. 

\subsection{Derivation of RG equations}
\label{subsec.rg}

For convenience, we rescale time and length to dimensionless units as $x_0=U\tau$ and $x_i=\Lambda r_i$ ($i=1,\ldots,D$), respectively, where 
$\Lambda$ is the initial momentum cut-off of the theory. Note, that $\Lambda$ is not sharply defined since we have performed a 
naive coarse graining in order to take the continuum limit and to invert the hopping matrix in the long-wavelength limit. In any case, one should expect $\Lambda=2\pi/a$ with 
$a$ of order unity or a few lattice constants. In the dimensionless units, the momentum cut-off is $\tilde{\Lambda}=1$. In addition, we rescale the fields as 
$\varphi_a=\psi_a\sqrt{K_3\Lambda^{2-D}/U}$, to obtain

\begin{eqnarray}
\S_\textrm{eff} & = & \S_0+\S_g+\S_h,\\
\S_0 & = & \sum_a\int_{\omega}\int_\bk \left(k^2-i\gamma_1\omega+\gamma_2\omega^2+r\right)|\tilde{\varphi}_a(\bk,\omega)|^2,\nn\\
\S_h & = & h\sum_a\int\ud x_0\int\ud^D\bx |\varphi_a(\bx,x_0)|^4,\nn\\ 
\S_g & = & g\sum_{ab}\int\ud x_0\int\ud x_0'\int\ud^D\bx |\varphi_a(\bx,x_0)|^2|\varphi_b(\bx,x_0')|^2,\nn
\label{Seffren}
\end{eqnarray}
with dimensionless coupling constants $\gamma_1=K_1U/(\Lambda^2 K_3)$ and $\gamma_2=K_2U^2/(\Lambda^2 K_3)$ for the temporal derivative terms, $\gamma_3=1$ for
the spatial gradient term, and $r=R/(\Lambda^2 K_3)$ for the mass coefficient. The rescaled interaction and disorder strengths are given by $h=HU/(\Lambda^{4-D}K_3^2)$ 
and $g=G/(\Lambda^{4-D}K_3^2)$. We wrote the bilinear actions $\S_0$ in frequency and momentum space using the Fourier transformation 
$\varphi_a(\bx,x_0) = \int_\bk\int_\omega
\tilde{\varphi}_a(\bk,\omega)e^{-i(\bk\bx+\omega\tau)}$.  For brevity,
we have defined $\int_\bk=(2\pi)^{-D}\int_{|\bk|\le1}\ud^D\bk$ and 
$\int_\omega=(2\pi)^{-1}\int_{-\infty}^\infty\ud\omega$. Note that since we focus on the zero-temperature limit, we can work with continuous frequency integrations rather than 
with discrete Matsubara summations.

\begin{figure}[t]
\includegraphics[width=0.95\linewidth]{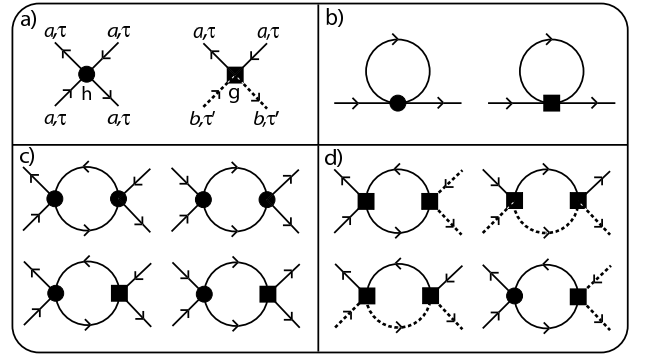}
\caption{a) Diagrammatic representation of the interaction vertex $h$ which is diagonal in the replica index and local in imaginary time and the replica mixing disorder vertex $g$,
which depends on two different imaginary times. b) Diagrams contributing to the renormalization of the mass coefficient $r$. 
Note that the contraction of the disorder vertex also gives rise to a renormalization of the temporal gradient terms $\gamma_1$ and $\gamma_2$ as described in the text. In c) and
d) the diagrams leading to a renormalization of $h$ and $g$, respectively, are shown. We restrict ourselves to one-loop order. Diagrams vanishing in the replica limit $n\to 0$ are not shown.}
\label{fig.diagrams}
\end{figure}

In order to derive the RG equations, we successively eliminate modes of highest energy with momenta on the infinitesimal shell $e^{-\ud l}\le |\bk|\le 1$ and 
consecutively rescale momenta, frequencies, and fields. To do so, we decompose the fields, 

\begin{eqnarray}
\varphi_a(\bx,x_0) & = & \varphi_a^<(\bx,x_0)+\varphi_a^>(\bx,x_0)\\
& = & \int_{|\bk|<e^{-\ud l}}\int_\omega \tilde{\varphi}_a(\bk,\omega)e^{-i(\bk\bx+\omega\tau)}\nn\\
& & +\int_{e^{-\ud l}<|\bk|<1}\int_\omega \tilde{\varphi}_a(\bk,\omega)e^{-i(\bk\bx+\omega\tau)},\nn
\end{eqnarray}
into 'fast' fields $\varphi_a^>$ containing modes from the infinitesimal momentum shell and 'slow' fields $\varphi_a^<$ depending on the remaining momenta. Decomposing
the action accordingly and taking the trace over the 'fast'
 fields $\varphi_a^>$, we find that the corrections of order $\ud l$ to the action $\S_\textrm{eff}^<=\S_0^<+\S_h^<+\S_g^<$ are 
given by $\ud\S_\textrm{eff}^<=\langle \S_h^{<>}+\S_g^{<>}\rangle_{0,>}-\frac 12 \langle (\S_h^{<>}+\S_g^{<>})^2\rangle_{0,>}^{(c)}$ where $\S_h^{<>}$ and $\S_g^{<>}$ 
contain all interaction and disorder vertices with two slow and two
fast fields. The superscript on the average over the quadratic terms
indicates that only connected diagrams contribute. The diagrams giving rise to a renormalization of order $\ud l$ of the coefficients of the bilinear action, as well as 
the interaction and disorder vertices are shown schematically in Fig. \ref{fig.diagrams}. Using the correlator

\begin{eqnarray}
\langle \varphi_a^*(\bk,\omega)\varphi_b(\bk',\omega')\rangle_0 & = & (2\pi)^{D+1}\delta_{ab}\delta(\bk-\bk')\nn\\
& & \times \delta(\omega-\omega')\C_0(k,r),\nn\\
\C_0(k,\omega) & = & (k^2-i\gamma_1\omega+\gamma_2\omega^2+r)^{-1},\quad
\label{corr}
\end{eqnarray}
we obtain immediately

\begin{equation}
\langle \S_h^{<>}\rangle_{0,>} = 4\frac{S_D}{(2\pi)^D} I_1 h \ud l  \sum_a\int_{\omega}\int_\bk^< |\tilde{\varphi}_a(\bk,\omega)|^2,
\label{intvertex}
\end{equation}
with $S_D$ the surface of the $D$-dimensional unit sphere and the inner frequency integral $I_1= \int_{-\infty}^\infty \frac{\ud\omega}{2\pi} \C_0(k=1,\omega)=
\int_{-\infty}^\infty \frac{\ud\omega}{2\pi} \C_0^*(k=1,\omega)=[4(1+r)\gamma_2+\gamma_1^2]^{-1/2}$ for 
$\gamma_2>0$ and $I_1=0$ for $\gamma_2<0$.  Comparing Eq. \ref{intvertex} with the bilinear part $\S_0$ of the effective action (\ref{Seffren}), one 
immediately recognizes that the contraction of the interaction vertex leads to a renormalization of the mass coefficient $r$. Due to the presence of two independent times
in the disorder vertex, the inner frequency of the 'fast' fields in the diagram  $\langle \S_g^{<>}\rangle_{0,>}$ is not free and we obtain 

\begin{equation}
\langle \S_g^{<>}\rangle_{0,>} = 2\frac{S_D}{(2\pi)^D} g \ud l  \sum_a\int_{\omega}\int_\bk^< \C_0(1,\omega)|\tilde{\varphi}_a(\bk,\omega)|^2.
\label{disvertex}
\end{equation}
Therefore, by expanding for small frequencies, $\C_0(1,\omega)\simeq I_0-I_0^2\gamma_1(-i\omega)-(I_0^2\gamma_2+I_0^3\gamma_1^2)\omega^2$ with 
$I_0=1/(1+r)$, we find that the disorder vertex not only renormalizes the mass but also the coefficients $\gamma_1$ and $\gamma_2$ of the temporal gradient terms. 

From the first-order vertex corrections $\langle \S_h^{<>}+\S_g^{<>}\rangle_{0,>}$ shown schematically in Fig. \ref{fig.diagrams}b, and from the rescaling of
frequency ($\omega\to \omega e^{d_z\ud l}$), momenta ($\bk\to \bk e^{\ud l}$), and fields ($\varphi_a\to\varphi_a e^{-\lambda\ud l}$), we obtain the renormalization-group 
equations for the coupling constants of the bilinear action,

\begin{subequations}
\begin{eqnarray}
\label{RGmass}
\frac{\ud r}{\ud l} & = & 2r + 2I_1 \overline{h}+I_0 \overline{g},\\
\frac{\ud \gamma_1}{\ud l} & = & -(d_z-2)\gamma_1-I_0^2\gamma_1\overline{g},\\
\frac{\ud \gamma_2}{\ud l} & = & -2(d_z-1)\gamma_2-I_0^2(I_0\gamma_1^2+\gamma_2)\overline{g},
\end{eqnarray} 
where we have set the scaling dimension of the fields to $\lambda=(D+d_z+2)/2$ to ensure that the coefficient $\gamma_3=1$ of the spatial gradient term is not renormalized.
Further, we have defined $\overline{h}=2S_D/(2\pi)^Dh$ and $\overline{g}=2S_D/(2\pi)^Dg$ to absorb pre-factors. 

The quadratic corrections $-\frac 12 \langle (\S_h^{<>}+\S_g^{<>})^2\rangle_{0,>}^{(c)}$ correspond to contractions of two vertices and lead to a renormalization of $h$ and 
$g$. The corresponding diagrams are shown in Fig. \ref{fig.diagrams}c
and  Fig. \ref{fig.diagrams}d, respectively, and are easily calculated
by using Eq. \ref{corr}.  Including the contributions from rescaling the RG equations for the interaction $\overline{h}$ and the disorder strength $\overline{g}$ are given by

\begin{eqnarray}
\frac{\ud \overline{h}}{\ud l}& = & -(D+d_z-4)\overline{h}-5I_2\overline{h}^2-6I_0^2\overline{g}\overline{h},\\
\frac{\ud \overline{g}}{\ud l}& = & -(D-4)\overline{g}-2(I_0^2+I_2)\overline{g}^2-4I_2\overline{g}\overline{h},
\end{eqnarray} 
\label{RGeqs}
\end{subequations}
where $I_2:=\int_{-\infty}^\infty\frac{\ud\omega}{2\pi} \C_0(1,\omega)\C_0^*(1,\omega)=I_0 I_1/2$ for $\gamma_2>0$ and $I_2=I_0/(2|\gamma_1|)$  for $\gamma_2<0$.

The above frequency integrals $I_1$ and $I_2$ are defined for $r+1>0$ which holds
in the MI phases as well as in the SF sufficiently close to the insulating states. Further, the RG equations are valid only for sufficiently
weak interaction $\overline{h}$ and disorder $\overline{g}$. Note that since the effective field theory is dual to the initial Bose-Hubbard model, the smallness of the interaction
vertex is guaranteed by $t/U\ll1$.

\subsection{Numerical Integration}
\label{subsec.numint}

In the following, we integrate the RG equations \ref{RGeqs}a-d numerically whereas the initial values of the coupling constants $\gamma_1(0)$, $\gamma_2(0)$, $r(0)$, 
$\overline{h}(0)$, and $\overline{g}(0)$ are determined by the microscopic parameters of the disordered BH model as explicitly derived in section \ref{sec.effective}. 
In the remainder of this section, we describe how the different phases of the disorder BH model 
are identified from the scale dependence of the coupling constants. The resulting phase diagrams are presented in section \ref{sec.results}.

Since $\gamma_1(0)\neq 0$ except for special values of $\overline{\mu}/U$ marking the tips of the Mott lobes, we use a dynamical exponent $d_z=2$. 
Although away from the tips, $\gamma_2$ eventually decreases exponentially under the RG, it is important to include $\gamma_2$ since the initial ratio 
$\gamma_2(0)/\gamma_1(0)$ becomes arbitrarily large in the vicinity of the tips, thereby strongly modifying the RG flow on small scales. 

Whereas on the mean-field level, the phase boundary is determined by the sign change of the bare mass coefficient $r(0)$ this parameter is renormalized by the 
interaction $\overline{h}$ and the effective disorder $\overline{g}$ leading to a shift of the phase boundary between the insulating and superfluid phases. 
In the insulating phases, $r(l)\to\infty$, corresponding to a freezing of the system on large scales, whereas in the SF phase $r(l)$ starts to diverge to negative infinite values. 
In the latter case, we stop the integration at a scale $l_s$ where $r(l_s)+1=0_+$ and the frequency integrals $I_i$ become singular. Note that the singularity
indicates that our approach is valid only in the strong coupling
regime. As soon as the system flows to weak coupling, the RG becomes unstable. However, sufficiently close
to the localized states it is possible to estimate the superfluid density as $\rho_s^2\sim -r(l_s)/\overline{h}(l_s)\exp(-Dl_s)$.

The instability of the MI towards the formation of a BG is indicated by the divergence of the effective disorder strength $\overline{g}$. Since the RG equations 
are valid only in the weak disorder regime, we stop integration at a scale $l^*$ where $|\overline{g}(l^*)|=1$ yielding an estimate $\xi\simeq a\exp(l^*)$ for the correlation length
corresponding to the typical linear dimension of Mott insulating islands. Note that the scaling dimension of $\overline{g}$ indicates the relevance of the effective disorder
in $D<4$ and hence in any real physical system. Therefore, even for an infinitesimally small initial value $\overline{g}(0)$, the system will eventually become unstable towards 
the formation of a BG although this might happen on astronomically large scales. In a finite system of linear dimension $L=a\exp(l_\textrm{max})$, the system 
looks ordered and is indistinguishable from an incompressible MI if $l_\textrm{max}<l^*$. In the following, we will investigate systems of different sizes, comparable to typical 
dimensions of optical lattices but also on the order of much larger
condensed matter systems, and determine the crossover between the MI
and the BG using
the condition $l_\textrm{max}=l^*$.

\section{Results}
\label{sec.results}

In the following, we determine the zero-temperature phase diagrams of the disordered BH model on a cubic lattice ($D=3$). For any set of parameters of the 
lattice model, chemical potential $\mu$, hopping $t$, on-site
repulsion $U$, and disorder parameters $\Delta$ and $\gamma$, we obtain the coupling constants of the effective 
field theory from the relations derived in Sec. \ref{sec.effective}. Note that the effective field theory provides a valid description of the long-wavelength physics only in the 
regime of  strong-coupling ($t/U\ll 1$) and weak disorder
($\Delta/U\ll 1$). By analyzing the RG flow of the
effective coupling constants, we are able to identify the different phases
delineated in Sec. \ref{sec.rg}.

\begin{figure}[t]
\includegraphics[width=\linewidth]{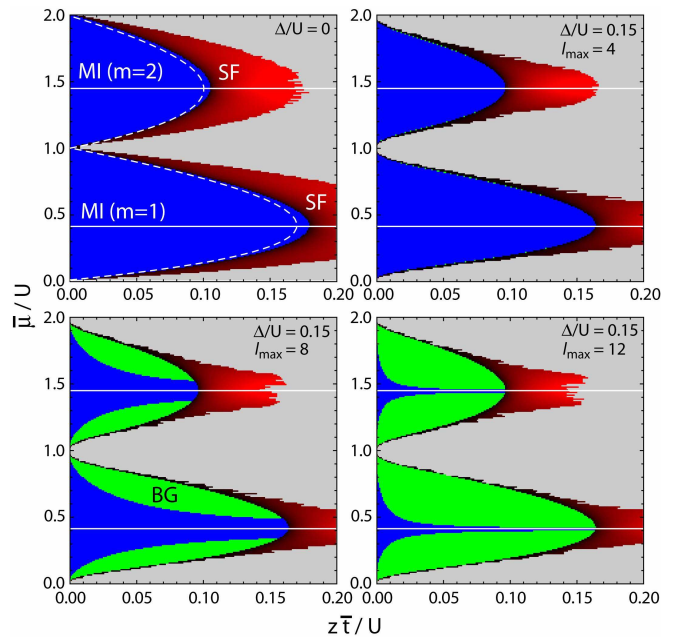}
\caption{(Color online) Upper left: Phase diagram for the clean system showing the $m=1$,  $m=2$ MI, and SF phases. The SF density $\rho_s$ is 
shown as a color gradient and, for comparison, the mean-field phase
boundaries are indicated by dashed white lines. Other panels show the
increase of the BG regions with the 
system size $L=a \exp(l_\textrm{max})$ for pure on-site disorder with $\Delta/U=0.15$ . Along the solid white lines, the effective disorder vanishes, $G=0$. 
Grey areas are not accessible within the present strong-coupling approach.}
\label{fig.size}
\end{figure}

As discussed in Sec. \ref{subsec.clean}, in the absence of disorder and at incommensurate fillings, the phase transitions from the MI states to the SF
are of mean-field type. However, although the interaction vertex $\overline{h}$ is irrelevant and decreases exponentially under the RG, it renormalizes the mass coefficient
$r$ to larger values as can be seen from Eq. \ref{RGmass}, leading to
a stabilization of the MI. In the upper left panel
of Fig. \ref{fig.size}, we show the increase 
of the Mott lobes due to the renormalization effect. 

We continue with the case of pure on-site disorder ($\gamma=0$). Already on a mean-field level, the presence of disorder leads to a destabilization of the Mott lobes since 
the bare value of the mass coefficient is reduced by disorder as can be seen from Eq. \ref{masscoeff} and $K_2^{(0)}>0$. In Fig. \ref{fig.size}, the phase diagrams 
obtained from the numerical integration of the RG equations for a disorder strength $\Delta/U=0.15$ are shown as a function of 
$\overline{\mu}/U$ and $z\overline{t}/U$. In addition to the decrease of the Mott lobes due to the mass reduction, disorder leads to an instability of the MI 
state towards the formation of a BG.  In Fig. \ref{fig.size}, the finite-size crossover between the two states for systems with different linear dimensions 
$L=a\exp(l_\textrm{max})$ is shown. Whereas for small systems comparable to typical optical lattice dimensions ($l_\textrm{max}=4$), the BG emerges only in a 
thin sliver close to the SF, the BG regions increase with the system size and almost completely take over the MI regions 
for $l_\textrm{max}=12$. This behavior is symptomatic of the fact that the BG phase is determined by exponentially rare compressible regions.\cite{Fisher+89}
While the BG regions increase with the system size, the different Mott regions become thinner and thinner and are centered around
$(\overline{\mu}/U)_m=\sqrt{m(m+1)}-1$ marking the tip positions of the Mott lobes in the clean system determined by $K_1^{(0)}=0$. Along these lines, the 
disorder vertex $G=-\frac 12 (K_1^{(0)})^2\Delta^2$ vanishes. Consequently, the tips are impervious to disorder (at least at one-loop order) and a direct transition 
from the MI to the SF obtains in the thermodynamic limit. 

\begin{figure}[t]
\includegraphics[width=\linewidth]{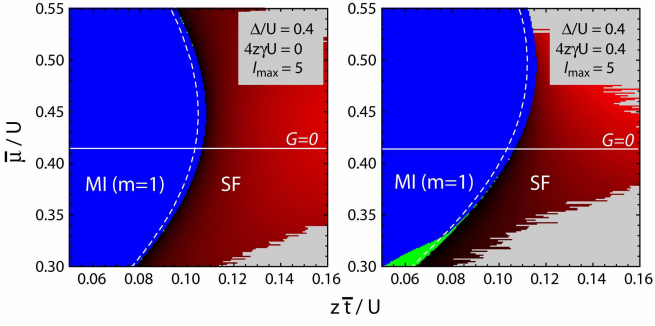}
\caption{(Color online) Enhancement and shift of the $m=1$ MI phase due to correlated hopping disorder (right) compared to the pure on-site disordered case (left).
The BG region is slightly enhanced due to the disorder in the hopping. Solid white lines indicate the vanishing of the effective disorder, dashed lines the corresponding 
mean-field (MF) phase boundaries determined by $R^{(0)}(\mu,t,U)=0$. For the BG phase boundary, no MF prediction is available.}
\label{fig.gamma}
\end{figure}

The effect of simultaneous correlated hopping disorder $\delta t_{ij}=\gamma(\epsilon_i-\epsilon_j)^2$ 
is illustrated in Fig. \ref{fig.gamma} for a speckle intensity of $\Delta/U=0.4$. Compared with pure on-site disorder ($\gamma=0$), already a small coupling 
$z\gamma U=0.1$ leads to a significant stabilization of the MI state as shown for the $m=1$ lobe in the close vicinity of the tip. Interestingly, the Mott lobes increase in 
size and simultaneously shift to larger values of the chemical potential $\overline{\mu}/U$. Further increase of $\gamma$ leads to further increase of the lobes whereas the 
effect is even stronger for larger fillings $m$. This behavior can be understood from inspecting the dependence of the mass coefficient $R$ on the 
disorder parameters (Eq. \ref{masscoeff}). Whereas potential disorder always leads to a decrease of the bare mass and therefore to a shrinkage of the Mott lobes, simultaneous 
correlated hopping disorder of the form studied here changes the bare mass coefficient by $4K_1^{(0)}\gamma\Delta^3/\overline{t}$. Recall that 
$K_1^{(0)}=-\partial R^{(0)}/\partial\mu$ changes sign at the tip
positions. For fillings below the tip of the Mott lobe, $K_1^{(0)}<0$, and therefore the Mott state is further destabilized. 
In contrast, above the tip, $K_1^{(0)}>0$ leading to an increase of the bare mass and hence to a stabilization of the Mott state as compared to pure on-site disorder. 
Although these arguments neglect renormalization effects and are based on the mean-field picture, the comparison with the phase diagrams obtained from numerical 
integration of the RG equations (see Fig. \ref{fig.gamma}) shows that renormalization
effects lead to an almost constant shift of the mean-field phase boundaries to
larger values of $z\overline{t}/U$, leaving intact the intuitive mean-field picture. 

\begin{figure}[t]
\includegraphics[width=0.95\linewidth]{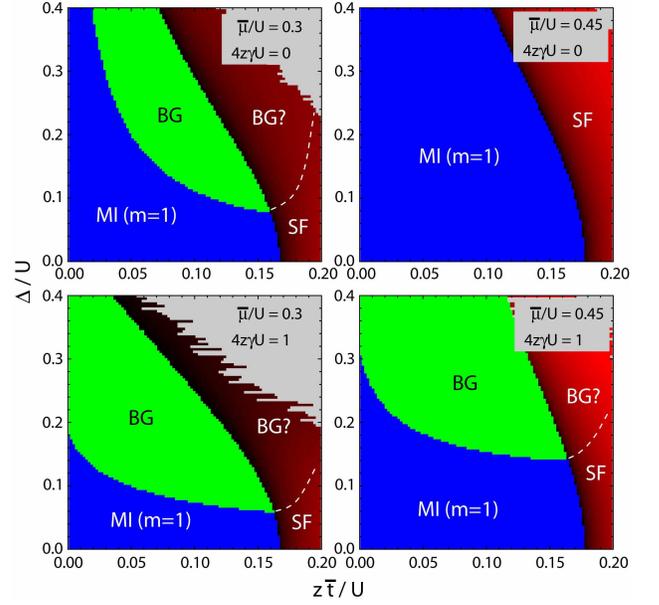}
\caption{(Color online) Phase diagrams as a function of speckle intensity $\Delta/U$ and relative hopping strength $z\overline{t}/U$ for values $\overline{\mu}/U=0.3$
(left panel) and $\overline{\mu}/U=0.45$ (right panel) of the chemical potential, corresponding to fillings below and slightly above the commensurate value $(\overline{\mu}/U)_1$, 
respectively. In the upper row, the phase diagrams for pure on-site disorder are shown and contrasted with the ones in the presence of simultaneous correlated hopping
disorder with coupling $4z\gamma U=1$ (lower row). The crossover between the MI and the BG is shown for a system size corresponding to $l_\textrm{max}=8$. Note that a reliable 
determination of the BG/SF transition is not possible within our approach. Alternative phase boundaries consistent with recent optical lattice experiments\cite{Pasienski+09} 
are indicated as dashed lines.}
\label{fig.Delta_t}
\end{figure}

Note that the phase diagrams in Fig. \ref{fig.gamma} are obtained by numerical integration up to a logarithmic scale $l_\textrm{max}=5$ corresponding to relatively 
small systems of roughly $150^3$ lattice sites. Whereas for pure on-site disorder the crossover from the MI to the BG sets in only very close to the superfluid phase, 
away from commensurate fillings (see also Fig. \ref{fig.size}), the
presence of simultaneous on-site and hopping disorder leads to a sizeable enhancement of the BG region 
as shown in Fig. \ref{fig.gamma} for $z\gamma U=0.1$. 
As in the case $\gamma=0$, the BG regions increase with the system size but never touch the lines at $(\overline{\mu}/U)_m$ corresponding to the 
Mott tips of the clean system.  As can be seen from Eq. \ref{dis}, the hopping-disorder contribution to $G$ is proportional 
to $K_1^{(0)}$ and therefore, regardless of the values of $\Delta$ and $\gamma$, the disorder vertex vanishes on approaching commensurate fillings.

To further investigate the evolution of the different phases as a function of the disorder strength and to further illustrate the effects of simultaneous correlated hopping 
disorder, we map out the phase diagrams as a function of the relative hopping strength $z\overline{t}/U$ and the speckle intensity $\Delta/U$ for different values of 
the chemical potential $\overline{\mu}/U$. In Fig. \ref{fig.Delta_t}, such phase diagrams are shown for $\overline{\mu}/U=0.3$, smaller than the value 
$(\overline{\mu}/U)_1=\sqrt{2}-1$ at the tip of the first Mott lobe and $\overline{\mu}/U=0.45$, slightly above the tip. For both values, the phase diagrams for pure on-site 
disorder are compared with the ones calculated in the presence of correlated hopping disorder with $4z\gamma U=1$, where the system size has been fixed 
as $l_\textrm{max}=8$.

Since the values of the chemical potential correspond to incommensurate fillings, the bare disorder vertex $\overline{g}(0)$ is different from zero and consequently, on a certain 
scale, $\overline{g}(l)$ will eventually become of order unity signaling the instability of the MI towards a BG. Therefore,  a direct MI-SF transition is impossible in the thermodynamic 
limit. In a finite system, the disorder strength $\Delta/U$ has  to exceed a certain threshold to induce a BG separating the MI from
the SF (see Fig. \ref{fig.Delta_t}). This critical disorder strength decreases with the system size but also depends on the value of $K_1^{(0)}$ and hence, on the value of $\overline{\mu}/U$. 
On approaching the commensurate value $(\overline{\mu}/U)_1$, $K_1^{(0)}$ decreases leading to 
an increase of the critical disorder strength necessary to induce a BG. This behavior can be clearly seen in Fig. \ref{fig.Delta_t}, both for pure on-site disorder and in the presence
of simultaneous hopping disorder. For $\overline{\mu}/U=0.3$ the critical disorder strength is much smaller than for $\overline{\mu}/U=0.45$ which is only slightly above 
the commensurate value. Note that in the latter case, for $\gamma=0$ the MI-SF remains direct up to the largest value $\Delta/U=0.4$ considered here.

The effect of simultaneous hopping disorder is in agreement with the trend suggested by Fig. \ref{fig.gamma}. For $\overline{\mu}/U=0.3$ below the commensurate value
$(\overline{\mu}/U)_1$ the phase transition to the SF is shifted to smaller values of $z\overline{t}/U$, whereas for $\overline{\mu}/U=0.45>(\overline{\mu}/U)_1$ the localized 
phases are stabilized by a finite $\gamma$. Moreover, it can be clearly seen that in the presence of correlated hopping disorder the system is more susceptible towards
Bose-glass formation. For all values $z\overline{t}/U$ in the MI regime the crossover to the BG sets in at smaller values $\Delta/U$ as compared to pure on-site disorder.

Interestingly, starting from the MI at $\overline{t}/U$ close to the SF phase, weak disorder can induce a transition to a SF rather than to a BG. 
Note that such a crossover is only possible in a finite system. As expected, sufficiently deep in the MI phase disorder induces the crossover to a BG. On further increase 
of the disorder strength we find a transition from a BG to a SF in agreement with phase diagrams obtained by stochastic mean-field theory\cite{Bissbort+09} and Quantum Monte
Carlo simulations in $D=3$,\cite{Pollet+09} but in disagreement with recent optical lattice experiments\cite{Pasienski+09} which seem to indicate that a disorder induced transition 
from an insulator to a SF is impossible. In fact, the transition between the BG and the SF can not be predicted within our approach. Since both the effective field theory and the RG 
analysis are valid only in regime of strong-coupling and weak disorder, we can only determine the instabilities of the MI, either towards the SF controlled by a flow to weak coupling, 
or towards the BG controlled by a flow to strong disorder. Alternative BG/SF phase boundaries which would be consistent with the recent optical lattice experiments\cite{Pasienski+09} 
are indicated by dashed lines in Fig. \ref{fig.Delta_t}. 

\begin{figure}[t]
\includegraphics[width=0.95\linewidth]{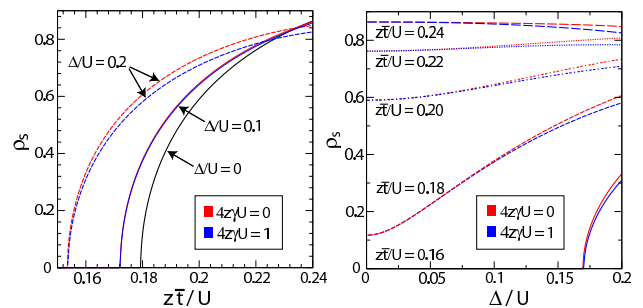}
\caption{(Color online) Evolution of the SF density as a function of $z\overline{t}/U$ (left) and $\Delta/U$ (right) for pure on-site disorder (red) 
and with correlated hopping disorder (blue).}
\label{fig.rhosf}
\end{figure}

Finally, we analyze the dependence of the SF density $\rho_s$ on the effective hopping strength $z\overline{t}/U$ and the speckle intensity $\Delta/U$ for pure on-site
disorder and in the presence of simultaneous hopping disorder. For simplicity, we focus on the point $(\overline{\mu}/U)_1$ marking the tip of the first Mott lobe in the clean 
system. In the panel on the left in Fig. \ref{fig.rhosf}, we show the onset and increase of $\rho_s$ as a function of $z\overline{t}/U$ for different values of $\Delta/U$. For larger 
values of $\Delta/U$, the transition shifts to smaller values of  $z\overline{t}/U$ signaling a shrinking of the insulating region. For all values of $\Delta/U$ the correlation 
$\gamma$ leads to a reduction of $\rho_s$ being more pronounced for
stronger speckle intensities. Whereas close to the transition,
$\rho_s$ increases with $\Delta/U$ (Fig. \ref{fig.rhosf}, right-hand panel) we find $\rho_s$ to decrease sufficiently deep in the SF. Simultaneous, correlated hopping disorder, inherent to an optical speckle, enhances this effect and moves the point where $\rho_s$ starts to decrease closer to the transition. 

\section{Summary and Discussion}
\label{sec.disc}
 
To summarize, we have studied the weakly disordered BH model on a cubic lattice in the strong-coupling regime and at zero temperature. By combining a disorder average
using the replica trick with a strong coupling expansion around the localized limit based on a Hubbard-Stratonovich transformation, we have derived the corresponding
long-wavelength field theory, providing explicit expressions relating the effective coupling constants to the microscopic parameters of the disordered BH model.
From the RG flow of the coupling constants, we have analyzed the instabilities of the MI state, either towards the formation of a SF characterized by a flow to weak coupling,
or towards a localized BG phase as indicated by a flow to strong disorder. 

The techniques presented here can be applied to various forms of simultaneous on-site and hopping disorder. 
Also the inclusion of finite temperature effects or an RG analysis beyond one-loop order are in principle possible.  In fact, the latter is necessary to capture the BG-SF 
transition at commensurate fillings.

Whereas the presence of simultaneous correlated hopping disorder as induced by optical speckle potentials does not change the overall topology of the phase diagrams, 
we have found that in comparison to pure on-site disorder ($\gamma=0$), the insulating phases are enhanced considerably, wheras the SF is destabilized. 
Moreover, the system becomes more susceptible towards the formation of a BG. 

Our phase diagrams are in qualitative agreement with stochastic mean-field theory\cite{Bissbort+09} 
and Quantum Monte Carlo simulations in $D=3$ in the weak disorder regime,\cite{Pollet+09} although the latter seem to disagree on the possibility of a finite-size crossover
between the MI and the SF in the presence of weak disorder. Unfortunately, at present, no complete experimental phase diagrams are available. 
In particular, it has not been possible yet to distinguish between the two insulating phases, the BG and the MI.\cite{Pasienski+09} In contradiction to our and 
previous\cite{Bissbort+09,Pollet+09} theoretical work the experiments seem to suggest that a disorder driven transition from an insulator (BG or MI) to a SF is not possible. 
However, a controlled calculation of the phase boundary between the BG and the SF is not possible within our present approach. 
The experimental observation that sufficiently deep in the SF phase the superfluid density decreases with the speckle intensity\cite{White+09} is in qualitative agreement 
with our predictions.  

One should bare in mind that a direct comparison of our theoretical phase diagrams  with the data obtained from optical lattice experiments suffers from
various potential problems. First of all, the typical dimension of optical lattice systems is much smaller than of realistic condensed matter system and therefore one should
expect much broader crossovers between different phases.  A second complication is the presence of an optical trap potential leading to inhomogeneous densities which 
might cause phase separation in the system or destroy commensuration effects which are crucial for understanding the BH model. Third, the experiments suffer from
heating by the lasers leading to temperatures which are not necessarily small compared to other experimental parameters as the effective Hubbard repulsion $U$. 
Also temperature changes during the experimental preparation and measuring processes are likely to obscure the data. 

However, despite all these challanges, the optical
lattice experiments have been billed as the ultimate quantum
simulators of paradigmatic lattice Hamiltonians such as the one defining the disorder BH model. In this work,
we have presented thermodynamic quantum phase diagrams for this model, where we have used disorder distributions and correlations as induced by an optical speckle lens
to arrive at a comprehensive picture.

Our major finding that the disorder vertex vanishes to leading order at commensurate boson 
fillings provides a plausible explanation for the long-standing controversy of wether a direct MI-to-SF is possible in the presence of weak disorder. On approaching
the tips of the Mott lobes, the bare value of the disorder vertex is strongly suppressed and consequently the scale on which the BG is observed becomes enormously large.
Not surprisingly, in earlier Quantum Monte Carlo (QMC)\cite{Scalettar+91,Krauth+91,Kisker+97,Sen+01,Lee+01} and density matrix renormalization group studies\cite{Pai+96} 
of relatively small systems, no indications of a BG have been found at or close to commensurate fillings whereas from more recent QMC data on larger two\cite{Prokofev+04} and 
three dimensional\cite{Pollet+09} systems it has been concluded that even at the tips, a small amount of disorder is sufficient to induce a BG disrupting the MI-SF transition.
However, based on numerical simulations it is impossible to answer the question if in the thermodynamic limit a direct MI-SF transition obtains in the presence of infinitesimally 
weak disorder. 

Various analytical methods\cite{Fisher+89,Mukhopadhyay+96,Freericks+96,Svistunov96,Herbut97,Herbut98,Weichman+08,Singh+92,Pazmandi+98,Bissbort+09} have been 
employed to find an answer to this question, arriving at different conclusions. While there is consensus on the absence of a direct MI-SF transition
at incommensurate fillings, with the exception of the stochastic mean-field theory\cite{Bissbort+09} which presumably gives wrong results below the upper critical dimension $D=4$, 
the case of commensurate boson fillings remains highly controversial. In their seminal paper, Fisher, et al.\cite{Fisher+89}, who raised the question in the first place, argued that 
a direct MI-SF transition at commensurate fillings was unlikely, though not fundamentally impossible. A work based on strong-coupling series expansions\cite{Freericks+96} 
arrived at the same conclusion based on the observation of Lifshitz rare regions. In the one-dimensional case, Svistunov\cite{Svistunov96} demonstrated by an RG argument
that commensuration is always asymptotically destroyed by disorder and hence that the transitions at commensurate and incommensurate fillings are controlled by the 
same BG fix-point. Based on a double epsilon expansion, Weichman and Mukhopadhyay\cite{Mukhopadhyay+96,Weichman+08} suggested the existence of a universal 
disordered BG fixed-point in higher dimensions, irrespective of the filling, at which particle-hole symmetry is statistically restored. Finally, in the limit of large fillings, Herbut\cite{Herbut97,Herbut98} demonstrated the absence of a MI-SF transition at commensurate fillings by using duality mappings to the sine-gordon model in $D=1$ and 
the three dimensional Higgs electrodynamics in $D=2$. Whereas all these works support the destruction of the direct MI-SF by infinitesimally weak disorder, using real-space RG techniques\cite{Singh+92,Pazmandi+98} a direct MI-SF transition at commensurate fillings up to a critical disorder strength is found.

In our present work, we find that at incommensurate fillings already an infinitesimal amount of disorder is sufficient to induce a BG separating the MI from the SF,
in agreement with previous results\cite{Fisher+89,Mukhopadhyay+96,Freericks+96,Svistunov96,Herbut97,Herbut98,Weichman+08,Singh+92,Pazmandi+98}. The vanishing of 
the disorder vertex $g$ at the tips of the Mott lobes to leading order $(\Delta/U)^2$ clearly indicates that commensuration effects are crucial for understanding how the system 
couples to an infinitesimal amount of disorder. 

Whereas on a one-loop level, consistently capturing disorder contributions of relative order $(\Delta/U)^2$, a direct MI-SF transition
obtains at commensurate fillings, this result is unlikely to hold at the level of a two-loops. First, the next order in the disorder cumulant expansion yields a contribution of order 
$(\Delta/U)^4$ to the disorder vertex $g$ which does not vanish at the tips. A contribution of the same order comes from the renormalization by the term 
$\sim \Delta^2[\varphi_a^*(\tau)\partial_\tau\varphi_a(\tau)][\varphi_b^*(\tau')\partial_\tau\varphi_b(\tau')]$ at one loop order,\cite{Mukhopadhyay+96,Weichman+08} which is less
relevant than the generic disorder vertex but has to be included at the tips as the leading order quartic disorder vertex. Such a term
arises in the replica theory from the coupling of potential disorder to the linear time derivative term. Therefore, $g$ does not remain
zero at the tips and the MI-SF is not protected by symmetry. However,
including this term in a one-loop calculation 
is no longer consistent as a term $\sim \Delta^2|\varphi_a(\tau)|^4|\varphi_b(\tau')|^4$, generated by the coupling of disorder to the interaction vertex, gives rise to a 
renormalization of $g$ of order $(\Delta/U)^2$ at two loop order.
Consequently, a two-loop calculation seems unavoidable to settle this issue.
 We believe that an extension of the calculations presented here in combination with the double epsilon expansion
by Weichman and Mukhopadhyay\cite{Mukhopadhyay+96,Weichman+08} allowing to identify potential disorder fixed points will provide conclusive answers, not only to the existence 
of the BG but also on the nature of the transitions.

\textbf{Acknowledgment:} We are grateful for stimulating discussions
with N. Prokofev, M. Troyer, P. Weichman, M. Fisher, M. White, B. DeMarco, S. Vishveshwara, E. Fradkin, 
and D. Ceperley. This work is supported by the 
NSF (DMR0605769)  and by the DARPA OLE program.

\end{document}